\documentclass{ws-rv961x669}
\pdfoutput=1 
\usepackage{ws-rv-van}     
\usepackage{ws-rv-thm}     
\usepackage{subfigure}     
\usepackage{wrapfig}
\makeindex

\begin{document}

\chapter[Fantasia of a Superfluid Universe]{Fantasia of a Superfluid Universe  \\                                                                                                                                                                                                                                          
                                                                                    -- In memory of Kerson Huang
}

\author[H-B. Low and C. Xiong]{Hwee-Boon Low and Chi Xiong}

\address{Institute of Advanced Studies,
\& School of Physical and Mathematical Sciences, \\
Nanyang Technological University, Singapore. \\
xiongchi@ntu.edu.sg }



\body


\hrulefill
\begin{wrapfigure}{r}{0.4\textwidth}
  \begin{center}
    \includegraphics[width=0.4\textwidth]{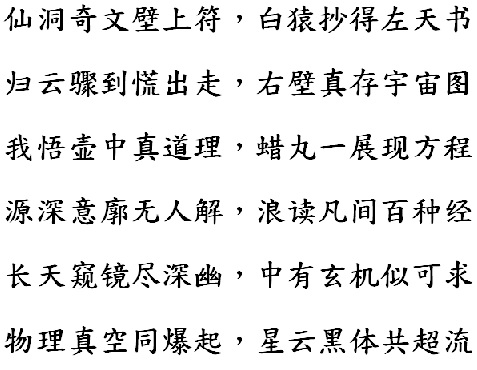}
  \end{center}
\end{wrapfigure}

{\it \small White Monkey stumbled into a sacred cave,

Copied the strange characters on the left wall,

And beat a hasty retreat, when clouds moved in,

Leaving the secret of Creation on the right wall.

I glimpsed at the truth in a wax-sealed slip,

Unravelled `n unfurled, revealing equations

Whose depth and scope seem to befuddle,

Learned scholars of classical citations.

Telescope peer deep into space,

Revealing God's great design, as angels knew it:

That Law and Vacuum rose together, and

Dark 'n bright matter co-move, in a superfluid.  

~ --- Epilogue from `A Superfluid Universe' \cite{SUbook}}

\hrulefill

When we first read this poem by Kerson Huang (1928-2016), we were puzzled on what is the ``strange characters"  that the white monkey copied from the left wall of the sacred cave, and what would be the ``secret of Creation on the right wall". Nobody had ever explained what is behind this ancient Chinese mythological story. Physicists nowadays, however, might provide an interpretation with the help of the Einstein equation   
\begin{equation}
R_{\mu\nu} - \frac{1}{2} g_{\mu\nu} R = T_{\mu\nu}
\end{equation}
where the ``left wall" has the Einstein tensor for gravity and the ``right wall" has the energy-momentum tensor for matter.   From the modern cosmological point of view, the creation of our universe and its evolution can be described by the Einstein equation. While the left-hand side of the equation is well-defined, the matter content on the right-hand side is still unknown, especially when dark matter and dark energy,  the``two dark clouds in the physics sky" of the 21st century ``moved in".

Kerson Huang ``glimpsed at the truth" and his theory on a superfluid universe consists of the following elements: I. choose the asymptotically free Halpern-Huang scalar field(s) to drive inflation; II. use quantum turbulence to create matter; III. consider dark energy as the energy density of the cosmic superfluid and dark matter the deviation of the superfluid density from its equilibrium value; IV. use quantum vorticity to explain phenomena such as the non-thermal filaments at the galactic center, the large voids in the galactic distribution, and the gravitational collapse of stars to fast-rotating blackholes. In the past few years this theory has been presented in Huang's articles and talks, as well as his recently published book \cite{SUbook}. We will briefly introduce this theory in the following sections.   

\section*{Macroscopic quantum phenomena and cosmology}

Macroscopic quantum phenomena, such as superfluidity, superconductivity and Bose-Einstein condensate, have attracted many theoretical and experimental researchers from different areas. Their relevance and applications in astrophysics and cosmology are particularly interesting and intriguing. For example, the idea of superconducting cosmic strings was pioneered by Witten \cite{Witten85} who proposed that cosmic strings can be turned into superconductors if electromagnetic gauge symmetry is spontaneously broken inside the strings. Almost at the same time, Zurek considered ``cosmological experiments in superfluid helium" and the analogy between cosmological strings and vortex lines in the superfluid \cite{Zurek85}. Later it has been shown \cite{Sin92}-\cite{Suarez} that Bose-Einstein condensates (BEC), which sustain superfluidity under certain conditions, can be formed in curved spacetime and considered as a candidate for dark matter.  In astrophysics superfluidity and superconductivity have been studied in neutron stars long time ago. Nevertheless, there are difficulties in generalizing these macroscopic quantum phenomena to physics at cosmic scales, and the most difficult one is probably its interdisciplinary nature, as mentioned in Huang's recent book \cite{SUbook}:

{\it ``It has now become difficult to be a generalist even in theoretic physics. You could be a particle theorist, cosmologist, or condensed matter theorist, but rarely all three! Even in particle theory, you would specialize in QCD, or electroweak theory, and in QCD you might specialize in confinement, lattice QCD, or quark-gluon plasma. Nature, however, does not know about our subdivisions, and exhibits phenomena that cut across all our specialities."}

One of the usual wrong impressions that people have about cosmic superfluidity, is the critical temperature for the superfluid phase transitions. Because of superfluid helium some people tend to relate superfluidity to low temperatures around $2$K, and do not realize that the critical temperature for superfluid phase transition depends on the system and hence, can be very high in certain systems. For instance, the surface temperature of a relatively old neutron star is at the order of $10^6$K.  

\section*{From order parameter to Higgs field}

Order parameters describe systems that change from a less ordered state to a more ordered one, or vice versa. How to choose order parameters depends on the system. In the case of BEC, the ``order" is related to a phase angle so the order parameter is chosen to be a complex field $\Psi$ satisfying the Gross-Pitaevskii (GP) equation
\begin{equation} \label{GPE}
i\hbar\frac{\partial\Psi}{\partial t}=\left(  -\frac{\hbar^{2}}{2m}\nabla^{2}
+\lambda \,|\Psi|^{2} - \mu_0 \right)  \Psi,
\end{equation} 
where $m$ is the atomic mass, $\mu_0$ the chemical potential. Eq. (\ref{GPE}) is a nonlinear Schr\"{o}dinger equation (NLSE) and not relativistic. It is not hard to find its relativistic generalization ($\Psi \rightarrow \phi$), in which the field $\phi$ satisfies a nonlinear Klein-Gordon (NLKG) equation derived from a Lagrangian density with a Higgs potential.  In terms of mathematical equations, the generalization from NLSE to NLKG is straightforward, while the physical meaning of the complex scalar field merits some attention. It can be considered as some Higgs field(s) spontaneously breaking certain global symmetry. However it is not clear what exactly this symmetry is. What we need is merely a U(1) subgroup embedded in some large symmetry group of the theory.   

We start with a general NLKG in curved spacetime and take the natural units $\hbar=c=1$. The vacuum complex scalar field is written in the polar form
$ \phi\left(  x\right)  =F\left(  x\right)  e^{i\sigma\left(  x\right)  } $. The action is
\begin{equation} \label{phiaction}
S=-\int dt\,d^{3}x\sqrt{-g}\left(  g^{\mu\nu}\partial_{\mu}\phi^{\ast}
\partial_{\nu}\phi+V\right)
\end{equation}
where $g^{\mu\nu}$ is the metric tensor, $g=$ det$\left(  g_{\mu\nu}\right)  $ and $V\left(  \phi^{\ast}\phi\right) $ is the self-interaction potential.  The equation of motion can be obtained 
\begin{equation} \label{NLKG}
\frac{1}{\sqrt{-g}}\partial_{\mu}\left(  \sqrt{-g}g^{\mu\nu}\partial_{\nu}
\phi\right)  -\frac{\partial V}{\partial\phi^{\ast}}=0
\end{equation}
In flat space-time, in the absence of galactic matter, the equation of motion reduces to an NLKG
\begin{equation}
\left(  \nabla^{2}-\frac{\partial^{2}}{\partial t^{2}}\right)  \phi
-\frac{\partial V}{\partial\phi^{\ast}}=0
\end{equation}
with a conserved current density
\begin{equation} \label{current}
j^{\mu}=\frac{1}{2i}\left(  \phi^{\ast}\partial^{\mu}\phi-\phi\partial^{\mu
}\phi^{\ast}\right)  =F^{2}\partial^{\mu}\sigma
\end{equation}
In the presence of a galaxy (a generic term that includes a star), we add to
the Lagrangian density an interaction term $\mathcal{L}_{\text{int}}$, which
represents the non-gravitational interaction between galaxy and scalar field.
(Gravity will be included later).
The galaxy is introduced as an external source, with a given energy current
density $J^{\mu}$. Its interaction with the scalar field is described
phenomenologically through a current-current interaction, as dictated by
Lorentz covariance:
\begin{equation} \label{Jmujmu}
\mathcal{L}_{\text{int}}=-\eta J^{\mu}j_{\mu}
\end{equation}
where $\eta$ is a coupling constant, and $J^{\mu}$ is a four-vector:
\begin{equation}
J^{\mu}=\left(  \rho,\mathbf{J}\right)
\end{equation}
and the given function $\rho\left(  x\right)  $ describes the density profile of the galaxy. 
Note that from (\ref{current}) and (\ref{Jmujmu}) we have $\mathcal{L}_{\text{int}} = -\eta F^2 J^{\mu} \partial_{\mu} \sigma$ which has, from particle physics point of view, the typical form of a coupling between the Noether current $J^{\mu}$ from the normal matter sector and the Nambu-Goldstone boson $\sigma$, if the U(1) symmetry is spontaneously broken. In fact $\eta F_0^2$ can be parametrized as $f_{\phi}^{-1}$ in analogy with the pion decay constant $f_{\pi}$.

\section*{Halpern-Huang potential and Inflation}

\begin{figure}
\centerline{\includegraphics[width=6.5cm]{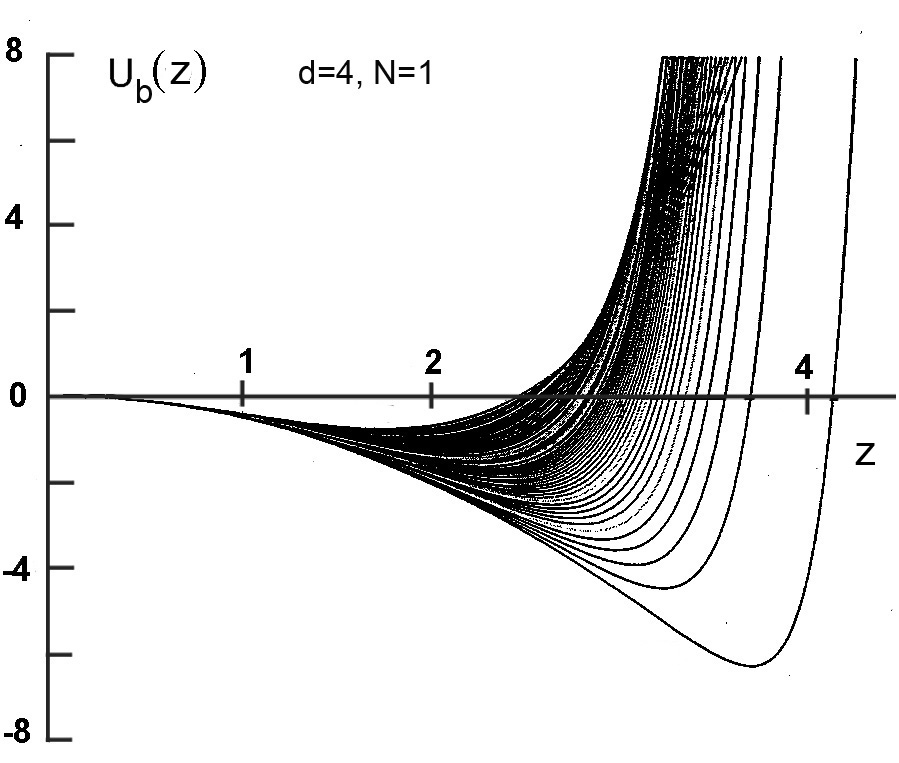}}
\caption{A family of Halpern-Huang potentials which increase exponentially at large field values.}
\label{HHP}
\end{figure}

A free massless scalar field is scale-invariant and corresponds to a Gaussian fixed point of the renormalization-group (RG) trajectory. If the inflaton is a scalar field emerging at the big bang from a Gaussian fixed point, then it should take a nontrivial direction to leave the fixed point. This suggests that the scalar field theory for the inflaton should be asymptotically free. The $\phi^4$ or a Higgs-type theory is certainly not asymptotically free, which motivated Huang et al to use a nontrivial scalar field theory for driving inflation \cite{HLT0, HLT1}. This leads to the Halpern-Huang potential \cite{HH}. It is generally believed that for renormalizable theories, only non-Abelian gauge theories can be asymptotically free. However, Huang and his student Halpern showed that an $N$-component scalar field theory with $O(N)$ symmetry and components $\varphi_1, \cdots, \varphi_N$ can be asymptotically free if the scalar potential has the form of \cite{HH}
\begin{equation} \label{hh}
U_b (z) = \frac{c}{\Lambda^b} \left[ \mathcal{M}\left( \frac{b-d}{d-2}, \frac{N}{2}, z \right) -1 \right], ~~~z = \frac{d-2}{2 \kappa} \sum^N_{n=1} \varphi^2_n
\end{equation} 
where $\mathcal{M}(p, q, z)$ is the Kummer function, $d$ the spacetime dimension and $b$ the eigenvalue of the linearized Polchinski equation (or the linearized exact RG-flow equation)
\begin{equation} \label{Pol}
\Lambda \frac{\partial U_b}{\partial \Lambda} = - b U_b.
\end{equation}
$b >0$ corresponds to asymptotic freedom. Fig. \ref{HHP} shows a family of Halpern-Huang potentials. Interestingly, the $d=2$ case reduces to the sine-Gordon theory. In $d=4$ the situation is complicated and a debate over the significance of the non-polynomial solutions of (\ref{Pol}) has continued for a long time (see e.g. a recent work \cite{Morris16} and references therein). To incorporate this scalar field theory into an inflation model, Huang et al suggested in ref. \cite{HLT0, HLT1} that the high momentum cutoff, $\Lambda$ must be related to the scalar factor $a(t)$ of the Robertson-Walker metric
\begin{equation}
\Lambda = \frac{\hbar}{a(t)}
\end{equation} 
and to obtain a set of self-consistent equations, one has to add to the pressure an extra term which turns out to be the trace anomaly. The main prediction of ref. \cite{HLT0, HLT1, HLT2} is that the Hubble parameter decays in time with a power law $H \sim t^{-p}$ and the universe expands at an accelerated rate $a \sim \exp t^{1-p}$ (see next section for the cosmological equations). 

\section*{Matter creation by vortex tangle}

In ref. \cite{HLT2} Huang et al generalized their inflation model by using a complex scalar field $\Phi = (\phi_1 + i \phi_2)/\sqrt{2} = F e^{i \sigma}$, and found that for matter-creation purpose, the complex scalar field must have emerged far from equilibrium in a state of quantum turbulence, i.e. a vortex tangle formed by vortex reconnections. Matter are efficiently created by the jets of kinetic energy produced in each reconnection, similar to how solar flares are produced in magnetic reconnections in the Sun. As discussed earlier, the order parameter of a superfluid or BEC should be promoted to a complex Higgs field in the relativistic regime, which satisfies an NLKG with a Halpern-Huang potential. 

Quantum vorticity is a manifestation of superfluidity. A quantum vortex is a circulating flow which satisfies the curl-free condition $\nabla \times {\bf v} = 0$ almost everywhere but still permits a net rotation with quantized circulation
\begin{equation}
\oint_C {\bf v} \cdot d{\bf r} =  \frac{\hbar}{m} \oint_C  \nabla \theta \cdot d{\bf r} =  \frac{\hbar}{m} ~2 \pi n, ~~n \in \mathbb{Z},
\end{equation}
where we have used the expression of a superflow velocity ${\bf v} = \hbar/m \nabla \theta$. Note that this is the non-relativistic relation and the order parameter is $\psi = |\psi| \exp(i \theta)$. The connection to the relativistic regime, e.g. the explicit relation between $\psi$ and $\Phi$ can be found in ref. \cite{XGGLH}.

The dynamics of vortices is extremely complex even without reconnections (think about the equation of motion of vortex lines and rings and their collisions). With reconnections, new vortex lines are formed when they cross one another and vortex rings break into smaller rings, eventually reaching a state called ``quantum turbulence" \cite{Feynman}. Quantum turbulence (QT) is a wide class of chaotic motions or tangled states of quantum vortices (vortex tangles) in quantum fluids \cite{Nemirovskii13}. Numerical simulations suggest that a vortex tangle has a fractal dimension of $\sim 1.6$ \cite{Kivotides01}. QT resembles classical turbulence (CT) in some aspects, e.g. the Kolmogorov energy spectrum in the inertial range and is different in other aspects, e.g. velocity distribution \cite{Lathrop08} in that CT has a Gaussian profile while QT has a power law tail coming from vortex reconnections which produce high-speed jets. It is this property of QT that Huang et al used for matter creation in early universe. More precisely, it has been proposed in ref. \cite{HLT2} that in the big bang era, a complex scalar field must have emerged far from equilibrium in a state of quantum turbulence or a vortex tangle with gigantic number of reconnections. Each reconnection produces two jets of kinetic energy for matter creation. It is then estimated that to produce the total mass in the present universe, the vortex tangle needed to last $\sim10^{-32} s$ \cite{HLT2}. A new and important ingredient in their cosmological equations is the inclusion of Vinen's equation which describes the energy density of the vortex tangle $\rho_v$, in addition to the energy density of a classical perfect fluid $\rho_m$ and the one from the regular part of the scalar field $\rho_\phi$
\begin{eqnarray} \label{HLT}
\dot{H} &=& \frac{k}{a^2} - (\rho + p) \cr
\rho &=& \rho_\phi + \rho_m + \rho_v, ~~~p = p_\phi + w_0 \, \rho_m \cr
\ddot{F} &=& - 3 H \dot{F} - F \langle v^2 \rangle - \frac{1}{2} \frac{\partial V}{\partial F} \cr
\dot{\rho}_v &=& - 3 H \rho_v + \alpha \rho^{3/2}_v - \beta \rho^2_v 
\end{eqnarray}   
plus a constraint $H^2 + k/a^2 - 2 \rho/3 = 0$ and $v^2$ is the energy density of induced superfluid flow, which contributes to the density $\rho_\phi$ and the pressure $\rho_\phi$ 
\begin{eqnarray}
\rho_\phi &=& \dot{F}^2 + V +  \langle v^2 \rangle \cr
p_\phi &=& \dot{F}^2 - V -  \langle v^2 \rangle - \frac{a}{3} \frac{\partial V}{\partial a}.
\end{eqnarray}
In Eq. (\ref{HLT}) the last line comes from the Vinen equation in RW background since $\rho_v$ is related to the vortex line density $l (t)$ as $\rho_v = \epsilon_0 l$ with $l$ satisfying
\begin{equation}
\dot{l} = - 3 H l + A l^{3/2} - B l^2.
\end{equation}
It is then shown that the equation group (\ref{HLT}) separates into two decoupled equation groups : the vortex-matter equation group and the scalar-cosmic expansion equation group. The former contains Vinen's equation governing the growth and decay of the vortex tangle, and an equation for the rate of matter creation. Its time scale is extremely small $(\mathcal{O} (10^{-18}))$ compared to the time scale used in the latter. This is because the vortex-matter system should be governed by a QCD energy scale of 1 GeV and the decoupling is possible due to the ``dimensional transmutation" \cite{HLT2}. 

\section*{Dark matter and galaxy collisions}

As mentioned earlier, the interesting idea of connecting dark matter and superfluidity has already been considered in refs. \cite{Sin92}-\cite{Suarez} and a lot of efforts have been devoted to considering dark matter as a Bose-Einstein condensate. This is because the standard Cold Dark Matter (CDM) paradigm may receive small-scale modifications to avoid the core-cusp problem, the angular momentum problem and etc., and macroscopic quantum phenomena such as superfluidity and BEC may be relevant as the temperature of the boson gas is below the critical temperature, BEC can form during the cosmological evolution of the universe. Most of these models require that the boson has an ultra-light mass of order $10^{-22}$eV. This corresponds to the typical size of the dark halos of galaxies ($ \sim$ kpc) and is totally different from the WIMPs  in the $\Lambda$CDM model which are much heavier (e.g. neutralino with a mass of the order of $10^2$GeV) and treated non-relativistically. The ultra-light feature of this dark matter boson motivated us in formulating a relativistic superfluid model with rotation terms  \cite{HXZ, XGGLH}, stressing on vorticity and its applications.        

\begin{figure}
\centerline{\includegraphics[width=7.5cm]{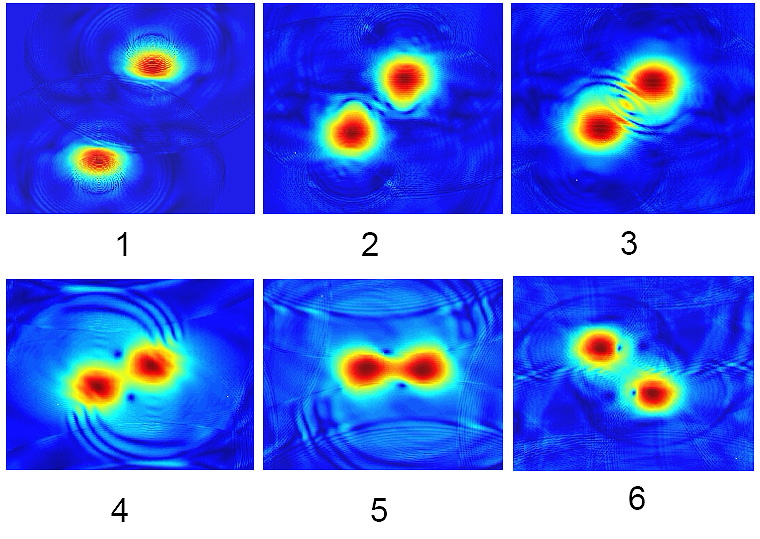}}
\caption{A simulation on galaxy collision in cosmic superfluid in (2+1)D space. The galactic halos move in accordance with superfluid hydrodynamics. The black dots are vortices produced by the shear motion between two galaxies.}
\label{galaxy}
\end{figure}

To describe the interaction between this dark matter boson and external sources such as the galaxies, Huang et al \cite{HLT1, HXZ, XGGLH} used the current-current coupling in Eq. (\ref{Jmujmu})
\begin{equation} 
\mathcal{L}_{\text{int}}=-\eta J^{\mu}j_{\mu} =  = -\eta F^2 J^{\mu} \partial_{\mu} \sigma.
\end{equation}  
For a rotating galaxy submerged in the cosmic superfluid with angular velocity ${\bf \Omega}$, the external current ${\bf J}$ can be written as ${\bf J} = \rho {\bf \Omega} \times {\bf r}$. We now include gravitational interactions in the Newtonian limit. The metric tensor is given by 
\begin{equation}
g_{00}    =-\left(  1+2U\right), ~~~~\left(  U  \ll 1\right)
\end{equation}
where $U$ is the gravitational potential. The equation of motion becomes
\begin{equation} \label{gNLKG}
-\left(  1-2U\right)  \frac{\partial^{2}\phi}{\partial t^{2}}+\nabla^{2}%
\phi+\frac{\partial U}{\partial t}\frac{\partial\phi}{\partial t}+\nabla
U\cdot\nabla\phi-\lambda\left(  \left\vert \phi\right\vert ^{2}-F_{0}%
^{2}\right)  \phi-i\eta\rho\left(  \frac{\partial\phi}{\partial t}%
+\mathbf{\Omega\times r}\cdot\nabla\phi\right)  =0
\end{equation}
The gravitational potential has contributions from the self-gravitation of the
scalar field, and from the galaxy's gravitational attraction:
\begin{align}
U\left(  x\right)   &  =-G\int d^{3}x^{\prime}\frac{\rho_{\text{sc}}\left(
x^{\prime}\right)  +\rho_{\text{galaxy}}\left(  x^{\prime}\right)
}{|x-x^{\prime}|}\nonumber\\
\rho_{\text{sc}}  &  =\left\vert \frac{\partial\phi}{\partial
t}\right\vert ^{2}+\left\vert \nabla\phi\right\vert ^{2}+\frac{\lambda}%
{2}\left(  \left\vert \phi\right\vert ^{2}-F_{0}^{2}\right)  ^{2}\nonumber\\
\rho_{\text{galaxy}}  &  =M_{\text{galaxy}}\rho
\end{align}
where $G$ is the gravitational constant and $M_{\text{galaxy}}$ is the galaxy's total mass. A galaxy immersed in the cosmic superfluid will become surrounded by a halo of higher superfluid density than the vacuum. This can be observed through gravitational lensing, and interpreted by us as dark matter. Based on (\ref{gNLKG}) we can simulate some galaxy dynamics. Fig. \ref{galaxy} shows two galaxies colliding at a nonvanishing impact parameter.  

It is worth noticing that the complex scalar might be a composite field, say a bound state or condensate of a fermion pair. For example in ref. \cite{Channuie} it has been proposed that the inflaton and the dark matter singlet can be considered as the scalar channel and the pseudoscalar channel, respectively, of a composite model based on the Nambu-Jona-Lasinio formalism. This might be related another superfluid ${}^3$He which could lead to another different superfluid universe scenario \cite{Volovik96, Volovik01}.

\section*{Rotations, vortices and Feynman's relation}

\begin{figure}
\centerline{\includegraphics[width=6.5cm]{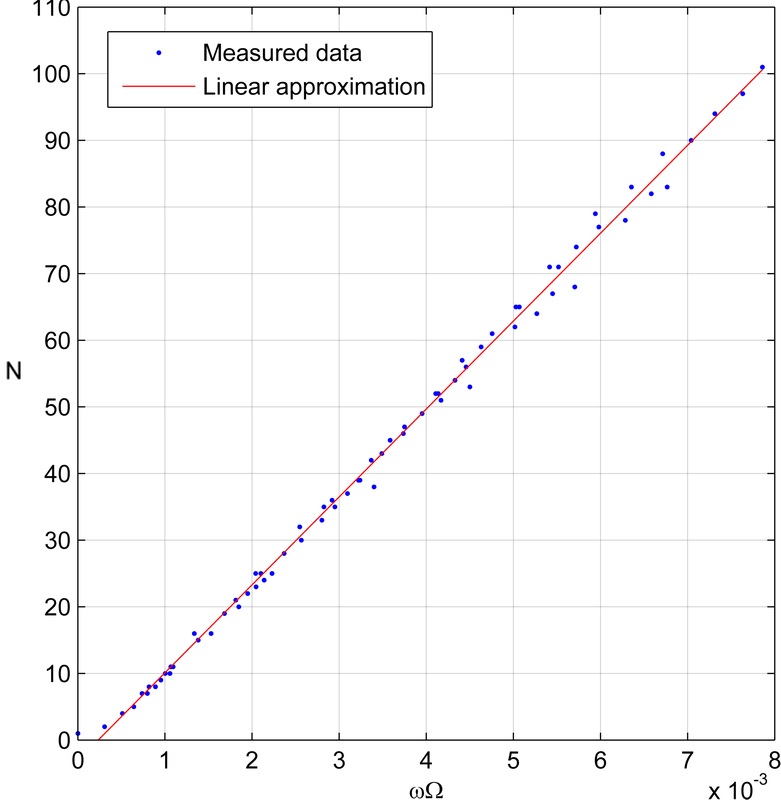}}
\caption{The number $N$ of vortices in a rotating bucket as function of the angular velocity $\Omega$ and the field frequency $\omega$. The straight line represents the relativistic Feynman relation (\ref{feynman}), which can be rewritten $N=(R^{2}/c^2)\omega\Omega$. Here, $R$ is the radius of the bucket, and $\omega$ is the frequency of the the stationary solution to the NLKG. }
\label{Feynman}
\end{figure}

In the large-scale motion of a quantum fluid, as in a classical medium such as the atmosphere or the ocean, vorticity is ubiquitous, being induced through different means in different systems. For superfluid helium, vortices can be created through rotation of the container, or through local heat perturbations. The latter can create quantum turbulence in the form of a vortex tangle \cite{Feynman, Donnelly, Nemirovskii13, Tsubota13}. Rotating BECs also produce vortices \cite{Fetter}. In the cosmos, quantized vortices in the background superfluid can be created by a rotating galaxy, colliding galaxies as shown in the previous section \cite{HXZ}, or a rotating black hole \cite{Good}. The big bang era could witness the creation of quantum turbulence. In ref. \cite{XGGLH} we have shown that all these effects can be traced to a universal mechanism characterized by operator terms in the NLKG that can be associated with the Coriolis and the centrifugal force:
\begin{align} \label{RR}
R_{\text{Coriolis}} &  =\frac{2\Omega}{c^{2}}\frac{\partial^{2}}{\partial
\phi\partial t},\nonumber\\
~~~R_{\text{centrifugal}} &  =-\frac{\Omega^{2}}{c^{2}}\frac{\partial^{2}%
}{\partial\phi^{2}},
\end{align}
where $\Omega$ is an angular velocity, and $\phi$ the angle of rotation. The angular velocity may be an externally given constant, or it may be a spacetime function arising from dynamics, interaction with external sources, or spacetime geometry.
\begin{figure}
\centerline{\includegraphics[width=7.5cm]{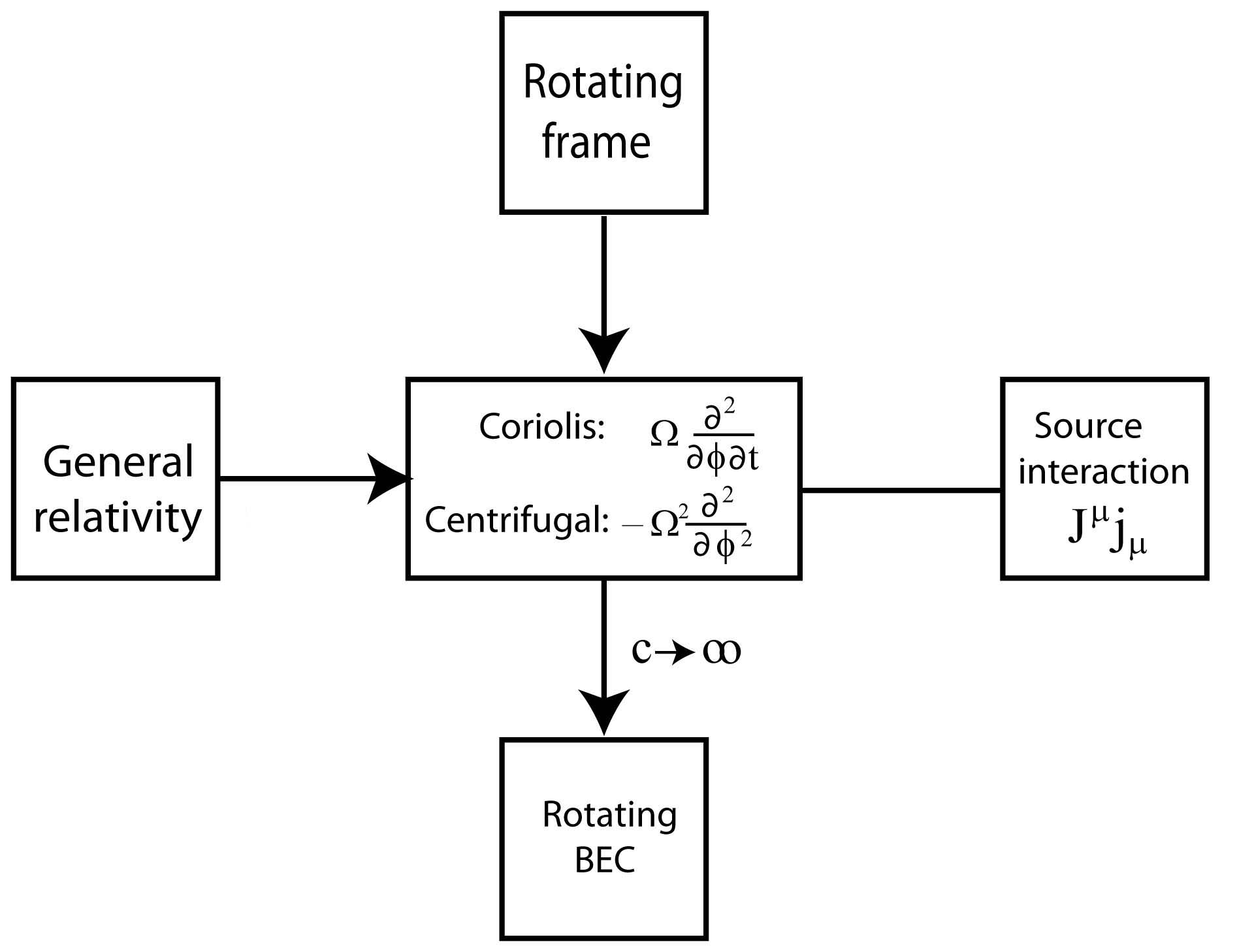}}
\caption{Chart showing various threads of the investigation in ref. \cite{XGGLH}. The general theme,
shown by the central and the left square, is that the mechanisms for vortex
creation are the inertia forces (Coriolis and centrifugal) in rotating
frames, generated by external means or by the spacetime metric, or by external
sources through a current-current interaction. (From ref. \cite{XGGLH})}
\label{chart}
\end{figure}
\begin{figure}
\centerline{\includegraphics[width=6.5cm]{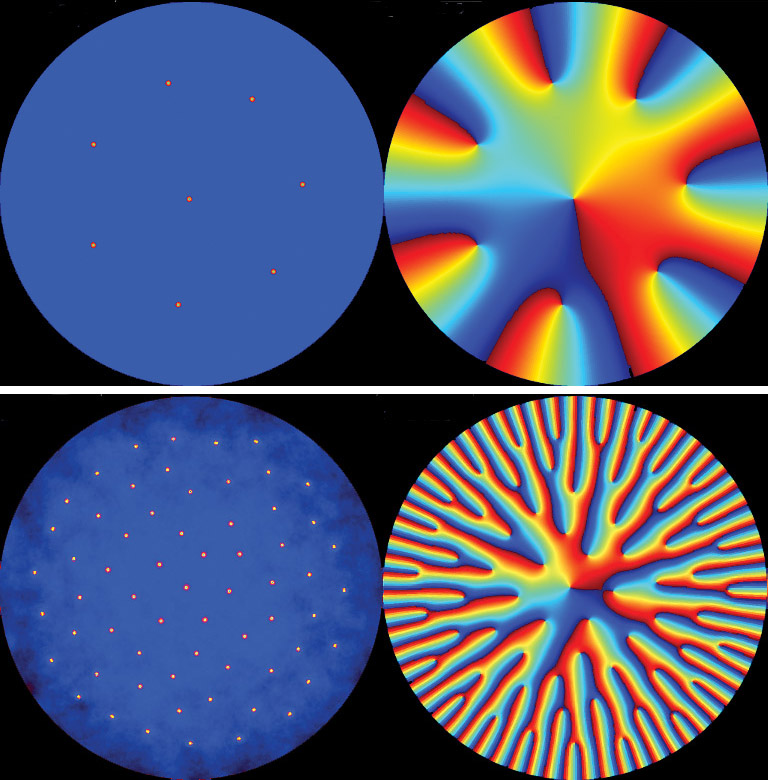}}
\caption{Two lattice states with 8 and 63 vortices respectively: The modulus of the field is plotted on the left side and its phase is on the right side. The color map from blue to red (see web version) indicates a phase change of $2\pi$. (From ref. \cite{XGGLH})}
\label{2VL}
\end{figure}
We have also generalized Feynman's relation to the relativistic regime \cite{XGGLH}. Consider $N$ vortices in rotating bucket of radius $R$ and angular velocity $\Omega.$ At the wall of the bucket, the superfluid velocity is $v_{\text{s}}=\Omega R$. Thus, $\nabla\sigma=v_{\text{s}}/\left(  c\xi_{\text{s}}\right)=\Omega R/\left(  c\xi_{\text{s}}\right)$ and note that $\xi_s = 1/\partial^0 \sigma$. From the relation ${\displaystyle\oint} ds\cdot\nabla\sigma=2\pi N$, we obtain
\begin{equation}
\Omega=\pi c\xi_{\text{s}}n_{\text{v}}, \label{feynman}
\end{equation}
where $n_{\text{v}}=N/\left( \pi R^{2}\right)  $ is the number of vortices per unit area. This formulas can give an estimate of the local vortex density when the superfluid flows with varying local angular velocity. In non-relativistic limit $\xi_{\text{s}}\rightarrow\hbar/mc$, we have $\Omega=\left(  \pi\hbar/m\right)  n_{\text{v}},$ which is called Feynman's relation. The derivation is valid only for large $N$, since we treat the rotation frame as if it were a rigid body. Fig.\ref{Feynman} shows in (2+1)-dimension the number of vortices $N$ in a rotating bucket as a function of  the product of the angular velocity $\Omega$ and the field frequency $\omega$.

Steady regularly-distributed vortex lattices are very interesting and distinctive macroscopic quantum phenomena, and one may wonder whether it can happen on astrophysical or cosmological scales. In ref. \cite{Good} it has been found that rotating black holes, such as the Kerr black hole in (3+1)-dimensions \cite{Kerr} and the BTZ black hole in (2+1)-dimensions \cite{Banados}, can provide a similar rotating environment as in condensed physics experiments and hence may induce quantum vortices in the cosmic superfluid or the Higgs vacuum. However, the rotating ``bucket'' here, significantly different from the container for liquid helium and the laser trap for BECs, is the spacetime itself.  

In general relativity, frame-dragging (or the Lense-Thirring effect) refers to a special distortion of spacetime geometry caused by a rotating mass. It occurs because the spacetime has angular momentum.  In ref. \cite{Good} it has been explored the possibility that the frame-dragging effect of a spacetime with angular momentum can create quantum vortices in the cosmic superfluid.  This is quantum vortex formation induced by angular momentum transfer from the spacetime to the field, and is referred to as {\it geometrical creation of quantized vorticity}.  

In general, this nonlinear Klein-Gordon equation should be solved in combination with the Einstein equation. Nevertheless, it is very difficult to find numerical solutions even for the cases neglecting the gravitational fluctuations. We are therefore compelled to make a slow-rotation approximation under which two rotation terms, a Coriolis term and a centrifugal term, emerge at first and second order in the rotational angular velocity $\Omega$, respectively, for both the BTZ case and the Kerr case. Taking the Kerr case as an example, we plug the Kerr metric in the NLKG (\ref{NLKG}) which yields
\begin{eqnarray}  \label{NLKG_Kerr}
\Sigma \, \square \Phi &=& \left[ -\frac{(r^2+\alpha^2)^2}{\Delta} +\alpha^2 \sin^2\theta \right] ~ \frac{\partial^2 \Phi}{\partial t^2} 
 - \frac{4 \alpha M r}{\Delta}  ~\frac{\partial^2 \Phi}{\partial t \partial \phi} 
 + \frac{\Sigma -2 M r }{\Delta \sin^2 \theta} ~\frac{\partial^2 \Phi}{\partial \phi^2}  \cr
&& + \frac{\partial}{\partial r} \left(\Delta \frac{\partial \Phi}{\partial r} \right) 
 + \frac{1}{\sin \theta}\frac{\partial}{\partial \theta} \left(\sin \theta \frac{\partial \Phi}{\partial \theta} \right)  \cr
&=& -  \Sigma \, [ \lambda(|\Phi|^{2}-F_{0}^{2})\Phi ],
\end{eqnarray}
where an overall factor of $\Sigma$ is included to simplify the expression. Similar to the BTZ case, we extract the Coriolis and centrifugal terms by expanding in powers of $\Omega_{\text{K}}$, to second order:
\begin{equation}
\label{Kerrrot}
R_{\text{Coriolis}}^{\text{K}}=\frac{2\Omega_{\text{K}}}{g_{tt}}\frac
{\partial^{2}\Phi}{\partial t\partial\phi},\text{ \ }R_{\text{centrifugal}%
}^{\text{K}}=\frac{\Omega_{\text{K}}^{2}}{g_{tt}}\frac{\partial^{2}\Phi
}{\partial\phi^{2}}.
\end{equation}
where $\Omega_{\text{K}}=- g_{t\phi} /g_{\phi\phi}$ is the local angular velocity. This is consistent with the results from a general study on NLKG in arbitrary rotational background \cite{XGGLH}, and in the flat spacetime limit it reduces to Eq. (\ref{RR}). 
Based on the numerical computations in refs. \cite{XGGLH, Good}, we find that the first order Coriolis term is the most important. We show that it leads to vortex formation, and therefore that quantum vortices can be generated by the frame-dragging effect of black holes. 

\section*{Vortex-ring lattice and the Kerson Layer}

\begin{figure}[ht]
\centerline{
\minifigure[A 2D vortex lattice surrounded by anti-vortices]
{\includegraphics[width=1.88in]{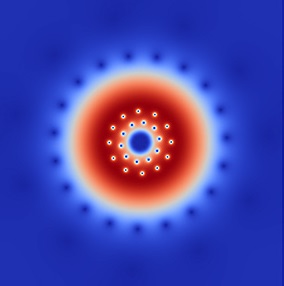}\label{cc_2d}}
\hspace*{4pt}
\minifigure[A 3D vortex-ring lattice]
{\includegraphics[width=2in]{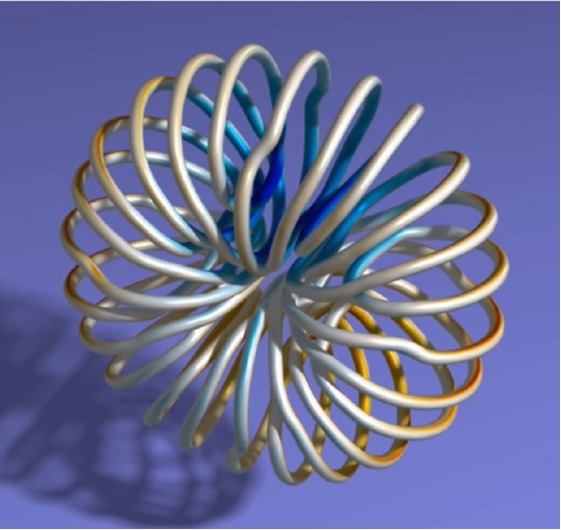}\label{cc_3d}}
}
\end{figure}

The current-current interaction (\ref{Jmujmu}) Huang introduced leads to numerical solutions of the NLKG which have, to our knowledge, never been found before. Fig.\ref{cc_2d} shows a solution in two-dimension with a vortex lattice surrounded by a circle of anti-vortices. This can be easily understood if we consider the two-dimensional case as a ``cross-section" of a three-dimensional vortex-ring lattice. Such solutions do exist as shown in Fig.\ref{cc_3d}. It resembles the magnetic flux tubes on the surface of the Sun, and demonstrates the similarity between the local rotation effect and the electromagnetic field.  Huang had been looking for this solution for a long time as he suggested that the so-called ``non-thermal filaments" observed near the center of the Milky Way (where there are super massive black holes) could be parts of vortex rings surrounding the blackholes, as pointed out in ref. \cite{SUbook, HLT2, XGGLH, Good}. In fact the current-current interaction (\ref{Jmujmu}) mimics rotation terms with local (space-dependent) angular velocities as shown above in the Kerr metric case, so it is possible to have a vortex-ring lattice as a solution to the NLKG in the Kerr metric.

For the gravitational collapse problem of rotating blackholes, the spacetime outside is described by the Kerr metric while the spacetime inside should be the generalization of the Robertson-Walker metric with angular momentum. This generalization, however, has not been well-established and hence, we do not have a gravitational collapse picture for most fast-rotating blackholes comparable to the Oppenheimer-Snyder solution \cite{OS}. Huang proposed \cite{SUbook, Good} that there may be a way out if a fast-rotating blackhole is immersed in a cosmic superfluid. The rotating blackholes can drag the neighboring superfluid into rotation by producing vortices. When the angular velocity is sufficiently high, the vortices may form a hydrodynamic boundary layer which can adjust to any boundary conditions required to connect the outside and the inside metrics. For this reason we call this boundary layer {\it Kerson Layer}.

\section*{Vortex dynamics and quantum turbulence}

\begin{figure}
\centerline{\includegraphics[width=6.5cm]{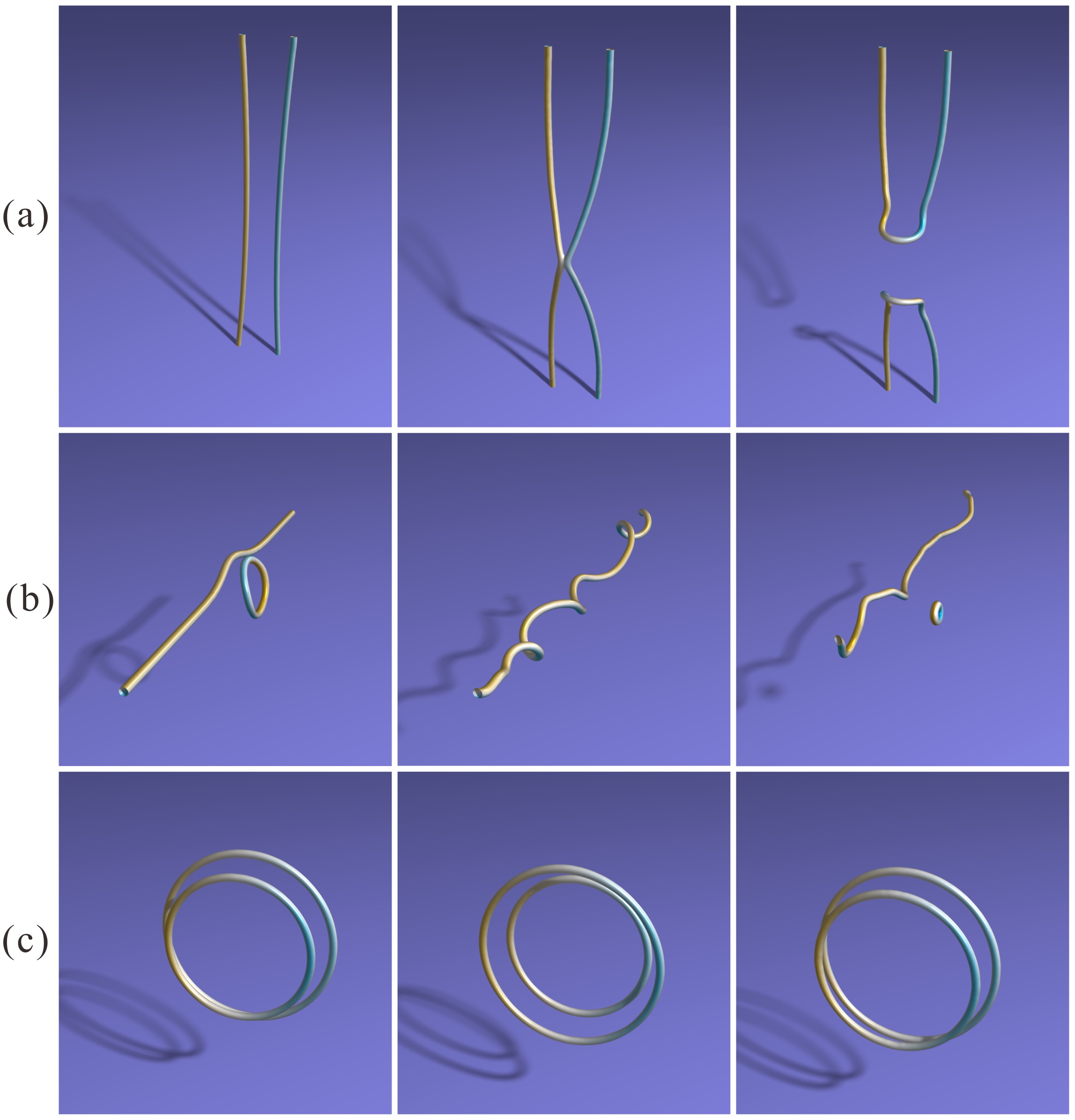}}
\caption{Dynamics of quantum vortices: (a) reconnection of two vortex lines; (b) scattering of a vortex ring on a vortex line; (c) leap-frogging of two vortex rings.}
\label{VD}
\end{figure}

\begin{figure}
\centerline{\includegraphics[width=6.5cm]{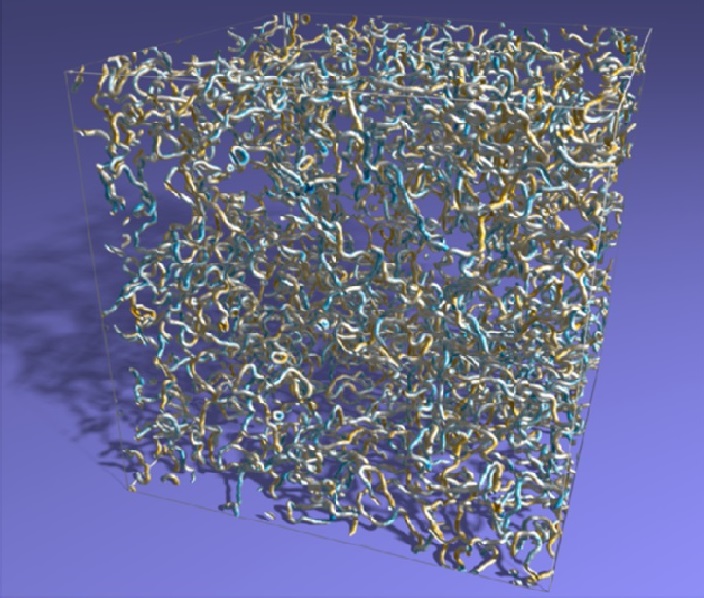}}
\caption{Quantum turbulence in the form of a vortex tangle.}
\label{QT}
\end{figure}

In classical fluid mechanics, vorticity and turbulence are notoriously complex topics while in quantum fluids quantized vortices and quantum turbulence, in the form of vortex tangle, are much simpler to study. Surprisingly, the vortex dynamics in superfluids can be described by the the nonlinear Schr\"{o}dinger equation (NLSE) or the nonlinear Klein-Gordon equation (NLKG). Kerson first suggested that we solve NLSE but later we decided to solve NLKG directly, since the latter can be reduced to the former as its non-relativistic limit. In ref. \cite{XGGLH} we solve numerically the NLKG with various inertial or interaction terms in (2+1)- and (3+1)-dimensional Minkowski spacetime. In Fig. \ref{VD} one can see how vortex lines cross and reconnect. The originally smooth vortex lines become lines with cusps at the reconnection point, and these cusps spring away from each other rapidly, creating two jets of energy in the superfluid. Through repeated reconnections, large vortex rings can be degraded until they become a vortex tangle as shown in Fig. \ref{QT}. This is the aforementioned mechanism for matter creation in the cosmos during the big bang era \cite{HLT1, HLT2}. For actual photographs of vortex reconnection in superfluid helium, see Ref. \cite{Lathrop}. More videos of our numerical computations and visualizations can be found in the Youtube channel ``IAS Computation Lab".

\section*{Galaxy formation, CMB and more}

Based on his early work with T. D. Lee and C. N. Yang on the hard-sphere model for Bose gases and superfluidity \cite{LHY1957}, the work with S. Weinberg on the ultimate temperature and thermodynamics in early universe \cite{Huang1970}, and the work with K. Halpern on the asymptotically-free scalar theory \cite{HH}, Kerson Huang wanted to incorporate all these elements into his theory of superfluid universe to study inflation, matter creation, dark matter and etc.  Huang was also interested in using quantum vorticity and turbulence to study galaxy formation, possible curl pattern in the cosmic microwave background (CMB) and etc. Given more time he should have been able to explore many unknown and interesting topics. However, Huang's own legend came to the last chapter and his book of life is closed on September 1, 2016. We could never forget how much we have learned from Huang, from his lectures, talks, discussions at lunch times and coffee breaks, and have always been admiring his insight, knowledge and devotion to physics. At his eighties, Huang was still working on algorithms, writing Matlab scripts to solve differential equations, and using Mathematica to calculate those Christoffel symbols and curvatures for the Kerr metric. In his book \cite{SUbook} Huang mentioned ``Whether or not the theory proves to be correct, I hope the reader will find this book interesting, perhaps even amusing." We dedicate this article to the memory of Kerson Huang, and wish that with this summary his theory would be appreciated by more researchers.   

\section*{Acknowledgements}
We thank M. Good, R-S Tung, X. Liu, Y. Guo, X. Zhao, A. Chua and H. Su for valuable discussions. We thank K. K. Phua and Guaning Su for their support without which this project would not have been possible.

\end{document}